\documentstyle[12pt]{article}
\begin{document}
\renewcommand{\theequation}{\thesection.\arabic{equation}}
\newcommand{\be}{\begin{eqnarray}}
\newcommand{\ee}{\end{eqnarray}}
\vskip 2.0cm
\begin{center}
{\Large\bf
Sphalerons, instantons, and standing waves on $S^3 \otimes R$.
}\\
\vskip 1.0cm

{\bf A.V. Smilga}\\
{\em Institute for Theoretical Physics, University of Bern, Sidlerstrasse 5,
CH-3012 Bern, Switzerland}
\footnote{Permanent address:
Institute for Theoretical and Experimental Physics, B. Cheremushkinskaya
25, Moscow 117259, Russia.}\\
\vskip 1cm
April 1995
\end{center}
\vskip 1.5cm
\begin{abstract}
We consider pure $SU(2)$ Yang-Mills theory when the space is compactified to
a 3-dimensional sphere with finite radius. The Euclidean classical self-dual
solutions of the equations of motion (the instantons) and the static finite
energy solutions (the sphalerons) which have been found earlier
are rewritten in handy physical variables with the gauge condition $A_0 = 0$.
Stationary solutions to the equations of motion in the Minkowski space-time
(the standing waves) are discussed. We briefly discuss also the theory defined
in a flat finite spherical box with rigid boundary conditions and present the
numerical solution describing the sphaleron.
\end{abstract}

\section{Introduction}
Topologically nontrivial nonabelian gauge field configurations characterized
by a non-zero Pontryagin number play a very important role in $QCD$. They
provide
the solution to $U(1)$ problem, are responsible for strong violation of the
Zweig rule in the scalar and pseudoscalar sectors and are crucial for
understanding the structure of $QCD$ vacuum state \cite{brmog}. Of all such
configurations, the distinguished position belongs to instantons \cite{BPST}
which are self-dual and present the solutions to the classical Euclidean
equations of motion. In a lot of papers, the role of instantons in $QCD$ was
studied numerically on lattices \cite{lat}. However, the lattice approximation
of $QCD$ involves two technical parameters: i) a finite ultraviolet cutoff
and ii) a finite size of the box where the theory is defined. Both effects lead
to modification and distortion of instantons, and it is important to understand
what these distortions are to keep them under control. In this paper, we
address the second
problem --- the modification of the classical BPST solution due to finite
volume effects.

In the recent paper \cite{baalinst},  a numerical study of the instantons
living
on the torus with periodic boundary conditions in spatial directions has been
performed. Also the sphaleron --- the static finite energy solution to the
Yang-Mills equations of motion with one unstable mode has been found.
(Such solutions do not exist in the Yang-Mills theory defined in infinite
space due to conformal invariance of the classical theory \cite{Col}
. However, a finite size of spatial torus introduces infrared cutoff which
breaks
down conformal invariance, and sphalerons appear by the same token as they do
in
electroweak theory where infrared cutoff is provided by the vacuum Higgs
average. \cite{Klink} ).

Cubic geometry is complicated, and little hope exists to find the solutions not
only numerically, but analytically. It is instructive therefore to study the
same problem with the spherical geometry where the analytical solution
{\it can} be obtained . It is rather easy to write down the solution for the
instanton living on a four-dimensional Euclidean sphere: when the coordinates
$x^\mu$ describing the stereographic mapping $R^4 \rightarrow S^4$ are chosen,
it just coincides with the standard BPST solution. However,
in such a geometry where time is also compactified, the instanton loses its
transparent physical meaning of the tunneling trajectory connecting
topologically non-equivalent vacua \cite{JR}. The geometry where only space is
compactified on $S^3$ and Euclidean time $\tau$ extends from $-\infty$ to
$\infty$ is much
more  interesting from the physical viewpoint {\it and} involves nontrivial
finite volume effects.

Analytical solutions for the Yang-Mills equations in such a geometry have been
studied earlier in a lot of papers \cite{Hoso}-\cite{baalS3}.
Actually, any solution on $S^3 \otimes R$ corresponds to some solution in the
flat space due to conformal invariance of the Yang-Mills equations and the fact
that the metrics of
$S^3 \otimes R$ is reduced to the flat metrics by a conformal transformation :
\begin{itemize}
\item The metrics of the Euclidean $R^4$ space can be written as \cite{baalS3}
 \be
 \label{Econf}
dx_\mu^2  = e^{2\tau/R} [d\tau^2 + R^2 d\Omega_3^2]
  \ee
with
 \be
  \label{rtau}
x_\mu^2 = R^2e^{2\tau/R}
 \ee
($d\Omega_3^2$ \  is the metrics on the unit 3-dimensional sphere.)
\item The metrics of the Minkowski $R^4$ space can be written as \cite{Steif1}
  \be
  \label{Mconf}
dt^2 -  d\vec{x}^2 = \ R^2 \ \frac{d\eta^2 - [d\chi^2 + \sin^2 \chi (d\theta^2
+
 \sin^2 \theta d\phi^2)]}{(\cos \eta + \cos \chi)^2}
  \ee
where
 \be
  \label{etachi}
t + r = R \ \tan \frac {\eta + \chi}2 \nonumber \\
t- r = R \ \tan \frac {\eta - \chi}2
  \ee
\end{itemize}
The solutions describing the standing waves on $S^3 \otimes R$ with real
Minkowski time were found first as peculiar non-stationary solutions in the
flat Minkowski space-time describing ingoing and outgoing spherical waves
\cite{Lus,Hoze}. [we see from (\ref{etachi}) that the relation of flat space
variables $t,\ r$ to the time ($\eta$) and spatial ($\chi$) variables on $S^3
\otimes R$ is rather intricate].

The sphaleron (a static solution with one unstable mode) on
$S^3 \otimes R$ has been found first in \cite{Hoso}, and it corresponds to the
well-known meron flat-space solution \cite{meron}:
 \be
  \label{meron}
 A^a_\mu = \frac {\eta_{\mu\nu}^a x_\nu}{x^2}
  \ee
where
$\eta_{\mu\nu}^a$ are the standard 't Hooft notations \cite{Hooft} :
  \be
  \label{eta}
\eta_{ij}^a = \epsilon_{aij},\ \eta_{i0}^a = -\eta_{0i}^a  = \delta_{ai}
  \ee

This solution does not involve a scale parameter and  the differential form
$A^a_\mu(x) dx_\mu$ does not depend on  $|x|$. After performing the
conformal transformation (\ref{Econf}), it translates into the static $\tau$
-- independent solution on $S^3 \otimes R$. Its time component \ [
corresponding to the radial component of Eq.(\ref{meron})] is zero.

Multiplying Eq. (\ref{meron}) by $dx_\mu$, we easily get
  \be
  \label{Hosbal}
A^{a (meron)}_\mu (x) dx_\mu \equiv A^{a (sphal)}_i (y) dy_i =
-\eta_{\mu\nu}^a e_\mu de_\nu
  \ee
where $y_i$ are some coordinates on $S^3$ and
$e_\mu = x_\mu/\sqrt{x^2}$ is the unit four-dimensional vector. The RHS of
(\ref{Hosbal}) is nothing else as the Maurer-Cartan forms $\sigma^a$ on
$S^3$, and this is the way the sphaleron solution has been presented in
\cite{Hoso,baalS3}.

The instanton solutions in
the flat Euclidean space have been translated into $S^3 \otimes R$ language in
\cite{baalS3}.

There are two {\it raisons d'etre} for the present paper.
The first one is methodical. We discuss all the solutions on $S^3 \otimes R$
found earlier in a uniform way and from the same angle treating the
finite 3-dimensional spatial sphere where the theory is defined as a
particularly convenient from the analytical viewpoint way to regularize the
theory in the infrared --- much as the standard compactification on the torus
used in the lattice approach is convenient for purposes of numerical
computations. Thus, we will  be interested neither in the gravitational
\cite{Hoso}-\cite{Steif1} nor in the flat-space \cite{Lus,Hoze} aspects of the
problem.
Our approach is similar to that of \cite{baalS3}. However, Minkowski space
solutions were not discussed in that paper and also the Euclidean solutions for
instantons and sphaleron on
$S^3 \otimes R$ were written  in rather
non-standard variables and were not easily visualizable. A considerable
fraction of this paper is devoted to rewriting these solutions in a physical
and transparent form and analyzing their properties. We work consistently in
the gauge $A_0 = 0$ where the  physical interpretation of instantons is most
evident.

 The next section is devoted to fixing the notations. In Sect. 3, we
 present an alternative
(compared to Refs. \cite{Hoso}-\cite{baalS3} ) derivation for the
sphaleron solution on
$S^3$ .  Sect. 4 is devoted to the instanton of the maximal size where the
volume density of action is the same at all points of $S^3$. This instanton
goes
over from the topologically trivial vacuum $A_i = 0$ at $\tau = - \infty$ to a
nontrivial one $A_i = i\Omega(\vec{x}) \partial_i \Omega^\dagger (\vec{x})$
at $\tau = \infty$ (where $\Omega(\vec{x})$ is a particular function describing
a
nontrivial mapping $SU(2) \rightarrow S^3$). It passes through the sphaleron at
$\tau = 0$. In Sect. 5, we write down the solution for instantons of arbitrary
size
$\lambda$ in handy terms, find the gauge transformation bringing it to the
Hamiltonian gauge $A_0 = 0$, and explore the limit $\lambda \ll R$ where the
solution is reduced to the standard BPST instanton.
In Sect. 6, we discuss some Minkowski space solutions describing
nonlinear standing waves on $S^3$ and get free of charge the increment
of instability for the sphaleron.

 Sect. 7 presents the second part of the paper. We discuss a related but
different problem where the system is
defined in the flat spherical box with the size $R$ and rigid boundary
conditions
$A_i (r = R)\  = \ 0$. No analytical solution can be obtained in this case, but
, by solving a simple ordinary differential equation, we find the sphaleron
solution numerically, draw its profile and calculate its energy.

\section{Notations.}
Different authors use different conventions and, for clarity, we list here
our own. The Yang-Mills action on an Euclidean curved manifold is
 \be
 \label{S}
    S = \frac 1{4g_0^2} \int \sqrt{g}\ d^4x \ G^{\mu\nu, a} G_{\mu\nu}^a
 \ee
where $G_{\mu\nu}^a = \partial_\mu A_\nu^a - \partial_\nu A_\mu^a +
\epsilon^{abc} A_\mu^b A_\nu^c$, $g_0$ is the coupling constant, and $g$ is the
determinant of metrics .
The $SU(2)$ gauge transformation is
  \be
  \label{Om}
  A^\Omega_\mu = \Omega (A_\mu + i\partial_\mu) \Omega^\dagger
 \ee
where $A_\mu = A_\mu^at^a = A_\mu^a \sigma^a/2$, $\sigma^a$ are the
Pauli matrices.
 The topological charge is
  \be
  \label{nu}
\nu = \frac 1{32\pi^2} \int g \ d^4x \ G^{\mu\nu, a} \tilde{G}_{\mu\nu, a}
   \ee
where $\tilde{G}_{\mu\nu}^a = \frac 12 \epsilon_{\mu\nu\alpha\beta} G^{\alpha
\beta, a}$ and the convention $\epsilon_{1230} = 1$ is chosen. Instanton is the
field configuration with the minimal action belonging to the topological class
$\nu = 1$. It is self-dual $G_{\mu\nu}^a = \sqrt{g}\ \tilde{G}_{\mu\nu}^a$
which means
  \be
  \label{sdual}
  E^a_i = - B^a_i
  \ee
where $E^a_i = G_{0i}^a,\ \ B^a_i = \frac 12 \epsilon_{ijk}\  \sqrt{g}\
G^{jk, a}$. In the following, we shall be interested with the case when
metrics is static
: $g_{00} = 1, \ g_{0i} = 0$ and $g_{ij}$ is time-independent.

In the Hamiltonian approach which is the most natural one for  static
geometry, an important characteristic of a configuration
$A^a_i(\vec{x}, \tau)$ is the Chern-Simons number
  \be
  \label{Q}
     Q = - \frac{\epsilon_{ijk}}{16\pi^2} \int g\ d\vec{x} \left [ {\rm Tr}
\{G^{ij} A^k\} + \frac {2i}3 {\rm Tr}\{A^iA^jA^k\} \right ]
  \ee
so that
 \be
 \label{nuQ}
\nu\  = \ Q(\tau = \infty) - Q(\tau = -\infty)
  \ee
Chern-Simons number is invariant under static topologically trivial gauge
transformations.

\section{Sphaleron on $S^3$.}
\setcounter{equation}0
 We choose the metrics of $S^3$ in the form
\be
\label{ds2}
ds^2 = \frac{d\vec{x}^2}{[1 + (r/2R)^2]^2}
  \ee
It is just the stereographic projection so that $r = 0$ corresponds to the
southern pole of the sphere, $r = \infty$ to the north pole and $r = 2R$
--- to the equator. In most formulae we shall set $R = 1$. The dependence on
$R$ can be anytime restored on dimensional grounds.

Let us look for a non-trivial solution of Yang-Mills equations of motion
$A^a_i(\vec{x})$ which lives on $S^3$ and has a finite energy. We assume that
the solution has spherical symmetry so that one can decompose
  \be
  \label{decomp}
A_i^a(\vec{x}) = \frac{1-\phi_2}r \ \epsilon_{aib}n_b + \frac{\phi_1}r
(\delta_{ai} - n_a n_i) + A n_a n_i \nonumber \\
,\ n_i = x^i /r
  \ee
where $\phi_{1,2}$ and $A$ are some functions of $r$. This Anzatz
is similar to Witten's Anzatz \cite{Witten} , only we keep $A^a_0 = 0$. The
magnetic field is
  \be
  \label{Bai}
B^a_i = (1 + r^2/4) \left[ \frac{\phi_1' - A\phi_2}r \epsilon_{aib}n_b +
\frac{\phi_2' + A\phi_1}r
(\delta_{ai} - n_a n_i) - \frac{n_a n_i}{r^2} (1 - \phi_1^2 - \phi_2^2)
\right]
  \ee
(prime means differentiation over $r$). The energy functional is
  \be
  \label{E}
  E = \frac {4\pi}{g_0^2} \int_0^\infty dr (1 + r^2/4) \left\{ (\phi_1' -
A\phi_2)^2
+ (\phi_2' + A\phi_1)^2  + \frac 1{2r^2} (1 - \phi_1^2 - \phi_2^2)^2 \right \}
  \ee
  On this stage, it is very convenient to introduce the radial and angular
variables
  \be
  \label{roal}
 \left \{
\begin{array}{c} \phi_1 = \rho \sin \alpha \\ \phi_2 = \rho \cos \alpha
\end{array} \right.
  \ee
(but we shall not necessarily assume $\rho$ to be positive). Variation of
(\ref{E}) over $A(r)$ gives the equation
  \be
  \label{adal}
A = \alpha'
  \ee
If the condition (\ref{adal}) is satisfied, the functional (\ref{E}) depends
only on $\rho(r)$ :
 \be
 \label{Ero}
  E = \frac {4\pi}{g_0^2} \int_0^\infty dr (1 + r^2/4) \left [
(\rho')^2 + \frac {(1-\rho^2)^2}{2r^2} \right ]
  \ee
Variation of (\ref{Ero}) over $\rho(r)$ gives the second equation of motion
  \be
  \label{eqror}
\rho'' + \frac {r/2}{1 + r^2/4} \rho' + \frac {\rho(1-\rho^2)}{r^2}  = 0
  \ee
A change of variables $r = 2e^u$ is handy after which Eq.(\ref{eqror}) acquires
the form \cite{Hoze}
  \be
  \label{eqrou}
\rho_{uu} + \rho_u \tanh u + \rho(1 - \rho^2) = 0
  \ee
Were the term $\propto \rho_u$ absent, the equation would describe a
conservative
motion in the potential
 \be
 \label{Vro}
V(\rho) = -(1 - \rho^2)^2/4
  \ee
 When the term with first
derivative is present, the ``energy'' is not conserved. When $u < 0$, it is
pumped into the system and when $u > 0$, it is dissipated.

If we want the physical energy (\ref{Ero}) of the field to be finite, the
function $\rho$ should be equal to 1 or -1 at $r = 0$ and $r = \infty$
(i.e. at $u = \pm \infty$). There are only two solutions to the equations
(\ref{eqror},\ \ref{eqrou}) having this property:
  \be
  \label{rosol}
 \rho = \mp \tanh u = \pm \frac{1 - r^2/4}{1 + r^2/4}
  \ee
These solutions are physically identical as the overall change of the sign in
$\rho$ can be traded for an overall phase shift $\alpha \rightarrow \alpha +
\pi$. Let us, for definiteness, choose the upper sign in Eq. (\ref{rosol}).

Speaking of the function $\alpha(r)$, it is not rigidly fixed. The only
requirement is that the {\it differential form} $A^a_i dx^i$ which, in contrast
to the potential $A^a_i(\vec{x})$, has an invariant meaning is uniquely defined
at $r = 0$ and $r = \infty$. Such a condition implies that
$A^a_i $ tends to zero at infinity faster than $1/r$. In that case,
$\alpha(r)$ can be any smooth function satisfying the boundary conditions
  \be
  \label{alas}
\alpha(0) = 0, \ \ \alpha(\infty) = (2n +1) \pi
 \ee
with integer $n$. The simplest case $n=0$ corresponds to the sphaleron
configuration. The case $n = -1$ describes the antisphaleron, and other values
of $n$
correspond to ``multisphaleron'' solutions to be discussed a bit later.
Whence the asymptotics of $\alpha(r)$ is fixed, the solutions with different
$\alpha(r)$ are obtained from each other by a topologically trivial gauge
transformation
  \be
  \label{gauge}
A_i(\vec{x}) \rightarrow \Omega(\vec{x}) \left(A_i(\vec{x}) + i\partial_i
\right) \Omega^\dagger (\vec{x})
 \ee
where $ \Omega(\vec{x}) = \exp\{i\beta(r) n_a \tau_a\}$ with $\beta(0) =
\beta(\infty) = 1$. One can be easily convinced that the transformation
(\ref{gauge}) acts on a general spherically symmetric Anzatz (\ref{decomp}) as
  \be
  \label{trans}
 \left\{ \begin{array}{l}
\rho^\Omega (r) = \rho(r) \\
\alpha^\Omega (r) = \alpha (r) + 2 \beta (r) \\
A^\Omega (r) = A(r) + 2 \beta'(r)
\end{array} \right.
  \ee
Naturally, the equations of motion (\ref{adal}) and (\ref{eqror}) are invariant
under this transformation.

We shall see in the next section that the especially clever choice of the phase
is
 \be
 \label{alsph}
\alpha (r) =  2 \arctan \frac r2
  \ee
(such $\alpha(r)$ just coincides with the polar angle on
$S^3$). In this particular gauge, the sphaleron solution acquires the form
 \be
 \label{sph}
A^a_i = \frac 1{(1 + r^2/4)^2} \left[ r\ \epsilon_{aib} n_b + (1 - r^2/4)
(\delta_{ai} - n_a n_i) + (1 + r^2/4) n_a n_i \right]
\ee
The solution (\ref{sph}) could of course be obtained directly from
(\ref{Hosbal}) if expressing $e_\mu$ in stereographic coordinates
  \be
  \label{ester}
  \left\{ \begin{array}{l}
e_0 \ = \ \frac{1 - r^2/4}{1 + r^2/4}\\
e_i \ = \ \frac {rn_i}{1 + r^2/4}
\end{array} \right.
  \ee
where $n_i = x^i/r$ is the unit 3-dimensional vector.

Antisphaleron has the same form but the signs in Eq. (\ref{alsph}) and of the
second and the third term
in square brackets in Eq. (\ref{sph}) are reversed. It corresponds to choosing
the same form for the phase function as in Eq.(\ref{alsph}) but with the
opposite sign.
 The energy of the sphaleron is obtained by substituting the solution
(\ref{rosol}) in (\ref{Ero}) :
  \be
  \label{EgR}
 E = \frac{3\pi^2}{g_0^2R}
\ee
where we have restored the dimensional factor $R^{-1}$. The
volume energy density of the sphaleron
  \be
  \label{eps}
 \epsilon = \frac 1{4g_0^2} F^a_{ij} F^{ij, a} = \frac 3{2g_0^2 R^4}
 \ee
is constant on the sphere.

The configuration (\ref{sph}) is a saddle point of the energy functional and
has an unstable mode. In Sect. 6, we shall find the increment of instability
$\mu$ for this mode. The result is
  \be
  \label{incr}
 \mu = \frac {\sqrt{2}}{R}
 \ee
The sphaleron solution (\ref{sph}) has 3 zero modes corresponding to a global
gauge rotation $A^a_i \rightarrow O^{ab}A^b_i$ where $O^{ab}$ is an orthogonal
matrix. The rotated field has the same energy and all other physical
properties, but it loses an explicit spherical symmetry and cannot
be presented in the form (\ref{decomp}).

Finally, let us calculate the Chern-Simons number (\ref{Q})  on the sphaleron
configuration. For any $A^a_i$ presented in the form (\ref{decomp}),
  \be
  \label{Qint}
Q = \frac 1{2\pi} \int_0^\infty dr \left[ A(1 - \phi_1^2 - \phi_2^2) + \phi_2
\phi_1' - \phi_1 \phi_2' - \phi_1' \right]
  \ee
For the sphaleron (antisphaleron), it is just
  \be
  \label{Qres}
 Q = \frac 1 {2\pi} \int_0^\infty dr \ \alpha'(r) = \pm \frac 12
  \ee
The sphaleron field is obtained from the antisphaleron one by the topologically
non-trivial gauge transformation
  \be
  \label{vac}
\Omega_0 = \exp \{ 2i\ \arctan \frac r2 \ n^a \tau^a\}
  \ee
Applying this transformation once more, one can get a multisphaleron
solution with the Chern-Simons number $Q = 3/2$ etc. The function $\rho(r)$
 for all such solutions is the same as for the sphaleron, and the phase
function
for the solution with Chern-Simons number $Q = (2n+1)/2$ is
  \be
 \label{nsph}
\alpha_n (r) =  2(2n+1) \arctan \frac r2
  \ee
, $n = 0, \pm1, \ldots$

\section{Instantons of maximal size.}
 \setcounter{equation}0
The instanton solution on $S^3 \otimes R$ which has the maximal size and is
spread homogeneously over the sphere has been written down in Ref.
\cite{baalS3}
in the following form
  \be
  \label{ibaal}
\left\{ \begin{array}{l}
A_0^a \ = \ 0 \\
A_i^a dx^i \ = \ - \frac {2e^{2\tau}}{1 + e^{2\tau}}
\eta_{\mu\nu}^a e_\mu d e_\nu
\end{array} \right.
  \ee
Subsituting there the expressions (\ref{ester}) for $e_\mu$ in
stereographic coordinates, we get
 \be
 \label{inst}
A^a_i = \frac {2e^{2\tau}}{e^{2\tau} + 1} \frac 1{(1 + r^2/4)^2} \left[ r
\epsilon_{aib} n_b + (1 - r^2/4)
(\delta_{ai} - n_a n_i) + (1 + r^2/4) n_a n_i \right]
\ee
One can be convinced that the instanton (\ref{inst}) satisfies the self-duality
condition (\ref{sdual}) and has the topological charge (\ref{nu}) $\nu = 1$ and
the action $S = 8\pi^2/g_0^2$.

We see that the Euclidean time dependence comes as a common factor, and the
spatial structure is exactly the same as for the sphaleron solution in the
gauge (\ref{alsph}). The instanton (\ref{inst}) starts from
the trivial vacuum
$A^a_i = 0$ at $\tau = -\infty$ and passes through the sphaleron (\ref{sph}) at
$\tau = 0$. At $\tau \rightarrow \infty$, the solution (\ref{inst}) tends to a
pure
gauge $A_i = i\Omega_0 \partial_i \Omega_0^\dagger$ with the same
 $\Omega_0$ as in
Eq.(\ref{vac}) and corresponds to a topologically nontrivial vacuum with
Chern-Simons number
$Q = 1$. Certainly, the instanton can be written down in any other gauge by
applying a gauge transformation (\ref{gauge}). After that, however, the
dependence on $\tau$ and $r$ would not factor out so nicely and also, at
$\tau =
-\infty$ , the potential would not be just zero but a pure gauge [with
a topologically trivial $\Omega(\vec{x})$].

This solution has 4 obvious zero modes: one of them corresponding to shifting
the time variable $\tau \rightarrow \tau - \tau_0$ and 3 zero modes
corresponding to
global gauge rotations. There is also the fifth zero mode corresponding to
going over to a solution with not the maximal size (such solutions will be
described in the next section), but there are no zero modes corresponding to
spatial translations --- the volume action density is the same at all points of
the sphere, and the solution has no centre.

  \section{Instantons of arbitrary  size.}
 \setcounter{equation}0
The solution of the arbitrary size has also been found in \cite{baalS3}.
It was presented in the following form (we still customized it a little bit)
  \be
  \label{Baalb}
A_0 \ = \ -\frac{s\tilde{b}^a\sigma^a}{1 + s^2 + b^2 + 2sb_\mu e_\mu}
\hspace{3cm} \nonumber \\
A_i dx^i \ = \ -\ \frac{(s^2 + s b_\mu e_\mu)\sigma^a\ +\ s e^{acd} \tilde{b}^c
\sigma^d}
{1 + s^2 + b^2 + 2sb_\mu e_\mu} \eta^a_{\mu\nu} e_\mu de_\nu
  \ee
where $b_\mu$ is a four-dimensional vector, $\tilde{b}^a = \eta^a_{\mu\nu}
b_\mu e_\nu$, $s = \xi e^\tau$, and $\sigma^a$ are the Pauli matrices.

To write down the solution for the instanton centered at $r = 0, \ \tau = 0$
and
expressed in standard variables, we have to do the following: {\it i)}
choose $b_\mu = (\vec{0}, -b)$ , {\it ii)} choose $\xi = \sqrt{1 + b^2}$,
and {\it iii)} to perform the stereographic projection (\ref{ester}).
It is convenient also to introduce the variable $\lambda = 1/b$. When $\lambda$
is small, it has the meaning of the physical instanton size. After
some calculations, one gets the following result
  \be
  \label{instb}
A_0^a = - \frac {rn_a}{C_+ \kappa\ \cosh(\tau) -
 C_-}  \hspace{8cm} \nonumber \\
A^a_i = \frac {1}{C_+[C_+ \kappa \cosh(\tau) -  C_-]}
\left[ r e^\tau \kappa \ \epsilon_{aic} n_c +
(e^\tau \kappa C_- -  C_+)  (\delta_{ai} - n_i n_a) + \right.\nonumber \\
\left. (e^\tau \kappa C_+ -  C_-) n_i n_a \right]
  \ee
where $$\kappa = \sqrt{1 + \lambda^2},\ \ \ C_\pm = 1 \pm r^2/4$$
The potential (\ref{instb}) is zero at $\tau = -\infty$.
At $\tau = \infty$, it has exactly the same asymptotics
$A_0 \rightarrow 0, A_i \rightarrow i\Omega_0 \partial
\Omega_0^\dagger$, $\Omega_0(\vec{x})$ being given by Eq.(\ref{vac}), as the
instanton of maximal size described in the previous section. The field strength
is now
  \be
  \label{BE}
E^a_i = -B^a_i = \frac {\lambda^2}{[C_+\kappa
 \cosh(\tau) -  C_-]^2}
\left[ r \ \epsilon_{aib} n_b + C_-
(\delta_{ai} - n_a n_i) + C_+\ n_a n_i \right]
  \ee
and the volume density of the action is no longer homogeneous in space.

The
instanton (\ref{Baalb}), (\ref{instb}) has 8 zero modes as it should.
As earlier, there are
global gauge rotations and time translations, but now there are four additional
collective variables $b_\mu$ describing the instanton scale and the spatial
position. To get the antiinstanton solution, one has to substitute
$\bar{\eta}^a_{\mu\nu}$ for $\eta^a_{\mu\nu}$ in Eq.(\ref{Baalb}) or, in our
language, to change the sign of $A_0^a$ and of the second and the third tensor
structures for $A_i^a$.

Let us first look what happens in the limit $\lambda \ll 1$. It corresponds to
 the
case when the physical size of the instanton is much less that the radius of
the sphere. If one also assumes that $r, |\tau| \ll 1$ so that the
effects due to
finite curvature on sphere are not essential (the requirement $|\tau| \ll 1$
is also
necessary because, at $|\tau| \sim 1$, the instanton field or whatever is
left of
it spreads out over the whole
sphere irrespectively of what the value of $\lambda$ is, and the expansion in
$r$ fails), the solution (\ref{instb}) acquires the familiar flat space form
  \be
  \label{BPST}
 A^a_\mu = \frac {2\eta_{\mu\nu}^a x_\nu}{\tau^2 + r^2 + \lambda^2}
  \ee
It is instructive to transform the solution still further and go over into the
Hamiltonian gauge $A_0 = 0$. It is achieved via a gauge transformation
$\Omega = \exp\{i\beta(\tau,r) n^a \sigma^a\}$.
Zero component of the gauge field is
killed provided $2\dot{\beta} n^a + A_0^a \ =\  0$ and hence
  \be
  \label{beta}
\beta(\tau,r) = \frac{r}{\sqrt{ r^2 + \lambda^2 C_+^2}}
\left\{ \arctan  \frac{e^\tau \kappa C_+ -  C_-}
{\sqrt{ r^2 + \lambda^2 C_+^2 }} \ + \ \arctan \frac {C_-}
{\sqrt{r^2 + \lambda^2 C_+^2  }} \right\}
  \ee
The second term in the braces appears due to initial condition
$\beta(\tau = -\infty, r)  \ = \ 0$ which is to be imposed if we want to
preserve
the property $A_i^a(\tau = -\infty, \vec{x}) = 0$. In the flat space limit, the
expression for $\beta(\tau, r)$ is simplified \cite{Actor}
  \be
  \label{betafl}
\beta_{flat} (\tau, r) = \frac r{\sqrt{r^2 + \lambda^2}} \left( \arctan
\frac \tau{\sqrt{r^2 + \lambda^2}} + \frac {\pi}2 \right)
  \ee
The transformed field $A_i^a$ can be easily found using the radial-angular
decomposition (\ref{roal}) and the recipe (\ref{trans}). As the full expression
is rather cumbersome, we write it down explicitly only in the flat space limit
:
 \be
 \label{flin0}
\rho_{flat} (\tau, r) = \frac {\sqrt{[\tau^2 + (\lambda + r)^2][\tau^2 +
(\lambda -
r)^2]}}{\tau^2 + r^2 + \lambda^2}  \hspace{8cm} \nonumber \\
\alpha_{flat} (\tau, r) =  \frac \tau{|\tau|}
\arccos \frac {\tau^2 + \lambda^2 -  r^2}
{\sqrt{[\tau^2 + (\lambda + r)^2][\tau^2 + (\lambda -
r)^2]}}
+ \frac {2r} {\sqrt{r^2 + \lambda^2}} \left( \arctan
\frac \tau{\sqrt{r^2 + \lambda^2}}
+ \frac {\pi}2 \right)  \nonumber \\
A_{flat}(\tau, r) = \frac {2\lambda^2}{r^2 + \lambda^2} \left[
\frac \tau{\tau^2 + r^2 + \lambda^2} + \frac 1{\sqrt{r^2 + \lambda^2}} \left(
\arctan
\frac \tau{\sqrt{r^2 + \lambda^2}} + \frac {\pi}2 \right) \right]
\hspace{2.5cm}
\ee
 In the limit $\tau \rightarrow \infty$,
\be
 \label{asflin}
\rho_{flat}(\infty, r) = 1 \hspace{1.5cm} \nonumber \\
\alpha_{flat} (\infty, r) = \frac {2\pi r}{\sqrt{r^2 + \lambda^2}}
  \ee
The phase rapidly grows from $\alpha = 0$ at $r = 0$ up to $\alpha = 2\pi$
at $r \gg \lambda$. It is no surprise, of course, that the instanton field
written in the gauge $A^a_0 = 0$ describes in the limit
$\tau \rightarrow \infty$ the twisted vacuum with
 \be
 Q(\tau = \infty) = \frac 1{2\pi} [ \alpha(\infty, \infty) - \alpha(\infty, 0)]
=
1
  \ee
It is noteworthy, however, that the required rise of the phase was provided
exactly by the gauge transformation (\ref{betafl}) which killed the nonzero
$A_0^a$ of the BPST solution.

It is instructive also to look at the phase of the full solution on $S^3$
at $\tau = \infty$ in the gauge $A_0 = 0$. It has the form
  \be
  \alpha_{S^3} (\infty, r) = 4 \arctan \frac r2 + \frac {2r}{\sqrt{r^2 +
\lambda^2C_+^2}} \left( \frac \pi 2 + \arctan \frac {C_-}
{\sqrt{r^2 + \lambda^2C_+^2}}  \right)
 \ee
Again, when the physical size of the instanton $\lambda$ is small, the phase
reaches its asymptotic value $\alpha = 2\pi$ at $r \sim \lambda$ and stays
there [at small $r$ it just coincides with the flat space expression
(\ref{asflin})]. Thus in this gauge, the solution has the nice property that
the
integral (\ref{Q}) for the Chern-Simons number is saturated in a
localized spatial region $r \sim \lambda$
(it was not the case for the original form (\ref{instb}) where the phase
 $\alpha_\infty(r) = 4\arctan (r/2)$ reached its asymptotic value only at $r
\gg 1$ close to the north pole of the sphere irrespectively of the
instanton size).

The final comment concerns the slice $\tau = 0$ of the instanton solution.
For the
instanton of the maximal size (\ref{inst}), it was just the sphaleron. When
$b \neq 0$, it is not a solution to the static equations of motion anymore, but
it still has (in the gauge $A_0 = 0$ !) the Chern-Simons number $Q = 1/2$. The
energy of such a configuration is $\propto 1/(g_0^2 \lambda)$ when the size of
the instanton is small and tends to the sphaleron energy (\ref{EgR}) when $b$
tends to zero (and $\lambda$ to infinity). It is still interesting to look at
the field configuration at $\tau = 0$ in more details. When the size of the
instanton $\lambda$ is small, it has the simple form
  \be
 \label{t=0}
\rho_{flat} (0, r) = \frac {|\lambda^2 - r^2|}{r^2 + \lambda^2}  \hspace{2.5cm}
\nonumber \\
\alpha_{flat} (0, r) =  \pm \pi \theta (r-\lambda)
+ \frac {\pi r} {\sqrt{r^2 + \lambda^2}} \nonumber \\
A_{flat}(0, r) = \frac {\pi \lambda^2}{(r^2 + \lambda^2)^{3/2}} \hspace{2.1cm}
 \ee
where the sign of the first term in the second equation depends on whether the
limit $\tau \rightarrow +0$ or the limit $\tau \rightarrow -0$ is taken.
This expression is singular.
The phase $\alpha_{flat} (0,r)$ is not uniquely defined at all, and whatever
sign is chosen, it is discontinuous at the point  $r = \lambda$ . Also
 the derivative $\rho'_{flat} (0,r)$ is discontinuous at that point. But
these are not physical singularities and depend on the particular convention
(\ref{t=0}). The potential $A_i^a(\vec{x})$ is a smooth function which is best
seen if expressing it in terms of
  \be
 \label{smooth}
\tilde{\rho} (0, r) = \frac {\lambda^2 - r^2}{r^2 + \lambda^2}
\hspace{0.3cm} \nonumber \\
\tilde{\alpha} (0, r) =
 \frac {\pi r} {\sqrt{r^2 + \lambda^2}}
\ee
(cf. Eqs.(\ref{rosol}), (\ref{alsph}) for the sphaleron potential on $S^3$).

\section{Standing waves.}
 \setcounter{equation}0
The explicit form of the sphaleron solution (\ref{sph}) allows one to find
rather easily some classical solutions to the Yang-Mills equations of motion
not only in Euclidean but also in Minkowski space-time. Let us search for such
solutions in the form
 \be
 \label{Ansatz}
A^a_i(t, \vec{x}) \ = \ \frac {h(t)}{C_+^2} \left[ r\ \epsilon_{aib} n_b + C_-
(\delta_{ai} - n_a n_i) + C_+ n_a n_i \right]
\ee
where $t$ is  the Minkowskian physical time.
With this Ansatz, the problem is greatly simplified and the lagrangian
constrained on the class of fields (\ref{Ansatz}) reads
  \be
  \label{Eh}
L = \frac 1{2g_0^2} \int \sqrt{g}\ d\vec{x} (E^a_i E^{ai} - B^a_i B^{ai} )
= \frac {3\pi^2}{g_0^2} [\dot{h}^2 - h^2 (2-h)^2]
 \ee
This is just the lagrangian of anharmonic oscillator. The potential
  \be
  \label{Vh}
V(h) = \frac {3\pi^2}{g_0^2} h^2 (2-h)^2
 \ee
has the minima at $h = 0$ and $h = 2$ which correspond to the classical vacua
with $Q = 0$ and $Q = 1$ . It also has the local maximum at $h = 1$ which
is the sphaleron ($Q = 1/2$). For general $h$, the Chern-Simons number of the
configuration (\ref{Ansatz}) is
\be
  \label{Qh}
Q = \frac {3h^2 - h^3}4
  \ee
The equation of motion is
  \be
  \label{moth}
\ddot{h} + 2h(1-h)(2-h) = 0
  \ee
A general solution to this equation is given by elliptic functions
\footnote{The equation (\ref{moth}) and its solution (\ref{ell}) appeared
first in Ref. \cite{Lus} where Minkowskian Yang-Mills solutions in flat
geometry
were studied. But the independent variable there was not just time, but
a complicated function of $t$ and $r$ [see Eq. (\ref{etachi})], and the
solution was not stationary.}
  \be
  \label{ell}
h(t) = \left\{ \begin{array}{l}
1 \pm \sqrt{c+1}\ {\rm dn} \left( (t - t_0) \sqrt{c+1} \left| \frac{2c}{1+c}
\right) \right. ,\ \ 0 \leq c \leq 1\\
\qquad \\
1 + \sqrt{c+1}\ {\rm cn} \left( (t - t_0) \sqrt{2c} \left| \frac{1+c}{2c}
\right) \right. , \ \ c > 1
 \end{array} \right.
 \ee
in the notations of \cite{Abr}. The constant $c$ determines the energy of the
solution:
  \be
  \label{Ec}
 E = \frac{3\pi^2}{g_0^2 R} c^2
 \ee
(where we have restored again the factor $R^{-1}$).

There are two distinguished physically interesting cases. The first one
corresponds to small $c$ when the solution (\ref{ell}) describes linear
standing waves with small amplitude $|h| \ll 1$ or $|2-h| \ll 1$. The frequency
of the oscillations is
  \be
  \label{omega}
  \omega = \frac 2R
  \ee
For small $R \ll \Lambda_{QCD}$ (i.e. when $g_0^2(R) \ll 1$), the frequency
(\ref{omega}) can be interpreted as one of the lowest excitation energies
of the
quantum Yang-Mills hamiltonian (a ``glueball mass'' if you will).

Another interesting case corresponds to the ``particle'' standing on the top of
the barrier at $t = -\infty$ which then starts to roll down, say, to the left,
turns back at $t = 0$, and approaches the top again at $t \rightarrow \infty$.
The explicit form of such $h(t)$ [corresponding to the choice
$c = 1, \ t_0 = 0$ and the lower sign in Eq. (\ref{ell})] is \cite{Steif}
  \be
  \label{hlim}
h(t) = 1 - \frac {\sqrt{2}}{\cosh(t\sqrt{2} /R)}
  \ee
The value $\mu = \sqrt{2}/R$ determines the eigenvalue of the unstable mode (or
increment of instability) of
the sphaleron solution as has already been quoted in Eq.(\ref{incr}).
The same result has been obtained in \cite{baalS3} in a different , more
complicated way.

Note that, when the system starts to roll sown from the sphaleron saddle point,
it approaches again the {\it sphaleron} configuration at $t \rightarrow
\infty$, not the antisphaleron one as one could probably guess in advance
knowing that  the potential should be periodic in the Chern-Simons number $Q$.

The potential is periodic, indeed, when {\it all} degrees of freedom of the
field are taken into account. But the Ansatz (\ref{Ansatz}) breaks down the
symmetry with respect to large gauge transformations, the antisphaleron
configuration is now unreachable, and the potential (\ref{Vh}) is not periodic
but has the reflecting walls on the left and on the right.
 Certainly, we could choose right from the beginning another Ansatz based
on the antisphaleron configuration or on any other configuration with the phase
(\ref{nsph}) and the Chern-Simons number $Q = (2n+1)/2$ which is related to
(\ref{sph}) by a topologically
nontrivial gauge transformation. We would get then corresponding Minkowski
space solutions which are the gauge copies of those we found.

One should also understand that , will all probability, the system is not
stable with respect to small perturbations of initial data distorting the
particular spatial dependence of the field (\ref{Ansatz}) . If it started to
roll down from the sphaleron ``mountain pass'' not exactly along the steepest
descent road (\ref{hlim}) , it would miss the pass on the way back, would be
reflected again by the slope of a mountain nearby , and starts to wander
randomly in the functional space (cf. \cite{Zhora,Zhora1}).

\section{Sphalerons in the ball.}
 \setcounter{equation}0
Let us consider now a similar but a different problem where the 3-dimensional
metrics is flat, but the field is defined in a finite spherical box $D^3$ with
the rigid boundary conditions
\be
\label{ballbc}
A_i^a(r=R)\  = 0\
  \ee
(and the gauge condition $A_0({\bf x}) = 0$ is assumed).

The boundary conditions (\ref{ballbc}) are similar in spirit to the bag
boundary conditions but do not coincide with the latter. Boundary counditions
(\ref{ballbc}) imply that the electric field and the normal component of the
magnetic field vanish on the boundary whereas the bag boundary conditions are
the vanishing of the normal component of electric field and the tangential
component of the magnetic field \cite{bag}. The bag boundary conditions
describe the spherical cavity in a dual superconductor and correspond to the
physical picture of individual hadron in confining medium.

For us, however, boundary conditions are just the way to regularize the theory
in the infrared and, for physical purposes, the size of the spherical box
should be chosen much larger than a characteristic hadron size. As the main
point of interest for us here are classical solutions, the bag boundary
conditions are not convenient --- they do not allow for non-trivial sphaleron
solutions. But the boundary conditions (\ref{ballbc}) do.

We assume again the
spherically
symmetric form (\ref{decomp}) of the potential and use the
radial-angular representation (\ref{roal}). One of the equations of motion for
the static sphaleron field retains the form (\ref{adal}) while the equation of
motion (\ref{eqror}) for $\rho(r)$ carrying nontrivial dynamic information is
modified to
    \be
  \label{eqrofl}
\rho'' + \frac {\rho(1-\rho^2)}{r^2}  = 0
  \ee
In accordance with the general theorem of Ref. \cite{Col}, this equation does
not have nontrivial finite-energy solutions when space has no boundaries.
Really, one can change the variable $r = e^u$ after which the equation
(\ref{eqrofl}) acquires the form
  \be
  \label{eqroufl}
\rho_{uu} - \rho_u  + \rho(1 - \rho^2) = 0
  \ee
and can be interpreted as a non-conservative motion in the potential
(\ref{Vro}) where energy is constantly pumped into the system \cite{Hoze}.
 The motion is
unbounded, and if it started a little bit away from the unstable equilibrium
points $\rho = \pm 1, \ \rho_u = 0$, it ends up at  $\rho(u) = \pm \infty$
at a finite value of $u$ \cite{Zhora1}.

But in finite box, the sphaleron solutions appear. In the full analogy with
the $S^3$
case considered earlier, the boundary conditions $\phi_1(R) = \phi_2(R) = A(R)
= 0$ and the condition of smoothness of $A_i^a(\vec{x})$ at origin imply
  \be
  \label{ro0R}
\rho(0) = \pm 1, \ \ \rho(R) = \mp 1
  \ee
Let us choose for definiteness the upper sign and be interested with the
sphaleron rather than with antisphaleron or multisphalerons.
Then the phase $\alpha(r)$ can be an arbitrary function satisfying the
conditions
  \be
  \label{bcal}
\alpha(0) = 0, \ \ \alpha(R) = \pi, \ \ \alpha'(R) = 0
  \ee
The function $\rho(r)$ is determined from the equation (\ref{eqrofl})
with the boundary conditions (\ref{ro0R}). The equation (\ref{eqroufl})
has even a name, it is
a special case of the Duffing equation \cite{Korn}, and, for space
without bounds,
it was studied extensively in \cite{Zhora1}, but we are not aware of any
presentation of its solution into known functions. Thus, we solved it
numerically. In Fig. 1, we plotted the solution for $\rho(r)$.
\footnote{As an amusing fact, one can note that in the region $r \sim 0$, the
function $\rho(r)$ behaves as $1 - 4.00\ r^2$. The closeness of the coefficient
to an integer suggests that it is equal to 4 exactly and may mean that an
analytic solution to the equation (\ref{eqrofl}) still can be found (?)}
In Fig. 2, the radial dependence of the volume energy density
  \be
\label{epsi}
\epsilon(r) = \frac 1{2g_0^2} B_i^a B_i^a (r) = \frac {1}{g_0^2 r^2}
\left\{ [\rho'(r)]^2 + [1 - \rho^2(r)]^2/(2r^2) \right\}
  \ee
is plotted. The total energy is
  \be
  \label{Esphfl}
E_{sph} = 4\pi \int_0^R \epsilon(r) r^2 dr \ \approx \frac {6.7 \pi^2}{g_0^2 R}
  \ee
Note that $\epsilon(R) \neq 0$ which means that the field strength $B_i^a$
(in contrast to the potential $A_i^a$) is non-zero at the boundary.
This {\it is}
the reason why we did not try to look for numerical solutions for the
self-duality equations $E_i^a = -B_i^a$ with the boundary conditions
$A_i^a(r=R)\  = \ 0$ which would describe instantons on $D^3 \otimes R$. With
all probability, such solutions just do not exist. Really, in the gauge $A_0^a
= 0$, the electric field $E_i^a = \dot{A}_i^a$ should be zero on the boundary
and
cannot coincide with $-B_i^a$ which is nonzero (at least for $t = 0$ where the
instanton of maximal size is expected to pass through the sphaleron).

The situation is essentially the same as in the standard electroweak model
where the infrared regularization is provided by the Higgs
 expectation value $v$.
Sphalerons do exist there \cite{Klink}, but instantons do not --- the
(classical) action of the instanton with the size $\rho$ :
$$S(\rho) = \frac{8\pi^2}{g^2} + 4\pi^2\rho^2 v^2 $$
depends on $\rho$ \cite{Hooft}. The action takes the minimal value in the
limit $\rho \rightarrow 0$ which is never achieved for non-singular
configurations. Obviously, the same is true in our case. Instantons of small
sizes practically do not feel the boundary, and their action is almost
$8\pi^2/g^2$. But this limit is never achieved for smooth field configurations.
\section{Acknowledgements.}

I am indebted to A. Kudryavtsev and P. Minkowski for illuminating discussions,
to M. Volkov who pointed my attention to the papers \cite{Hoso}-\cite{Steif1},
to P. van Baal for useful correspondence, and to P. B\"uttiker who helped me
to run the Mathematica program. I acknowledge warm
hospitality extended during my stay at University of Bern where this work has
been done. This work was supported in part by Schweizerischer Nationalfonds and
by the INTAS grant 93-0283.

\section*{Figure captions.}
{\bf Fig. 1}. Numerical solution for the function $\rho(r)$ describing the
sphaleron in $D^3$.
Radius of the ball $R$ is set to 1. \\

\noindent {\bf Fig. 2}. Volume energy density of the sphaleron $\epsilon (r)$.
The normalization $R = g_0^2 = 1$ is chosen.

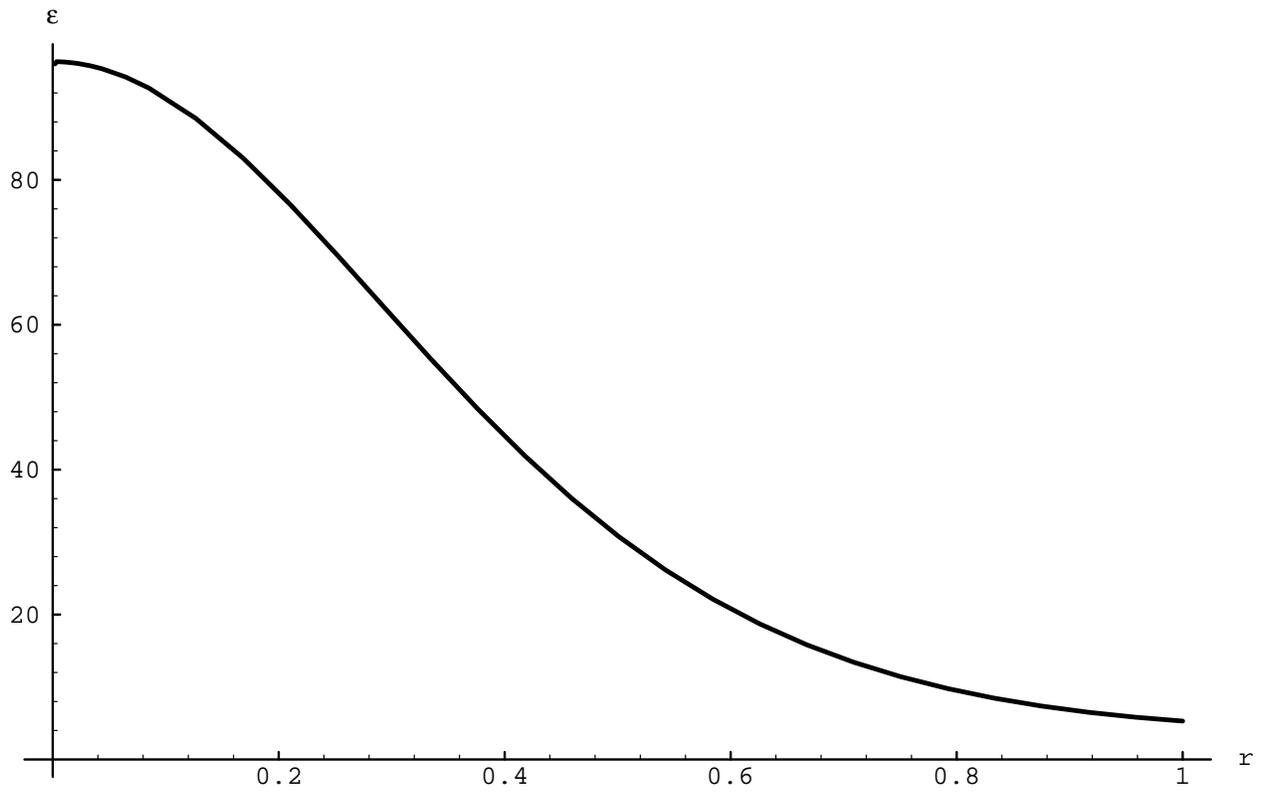
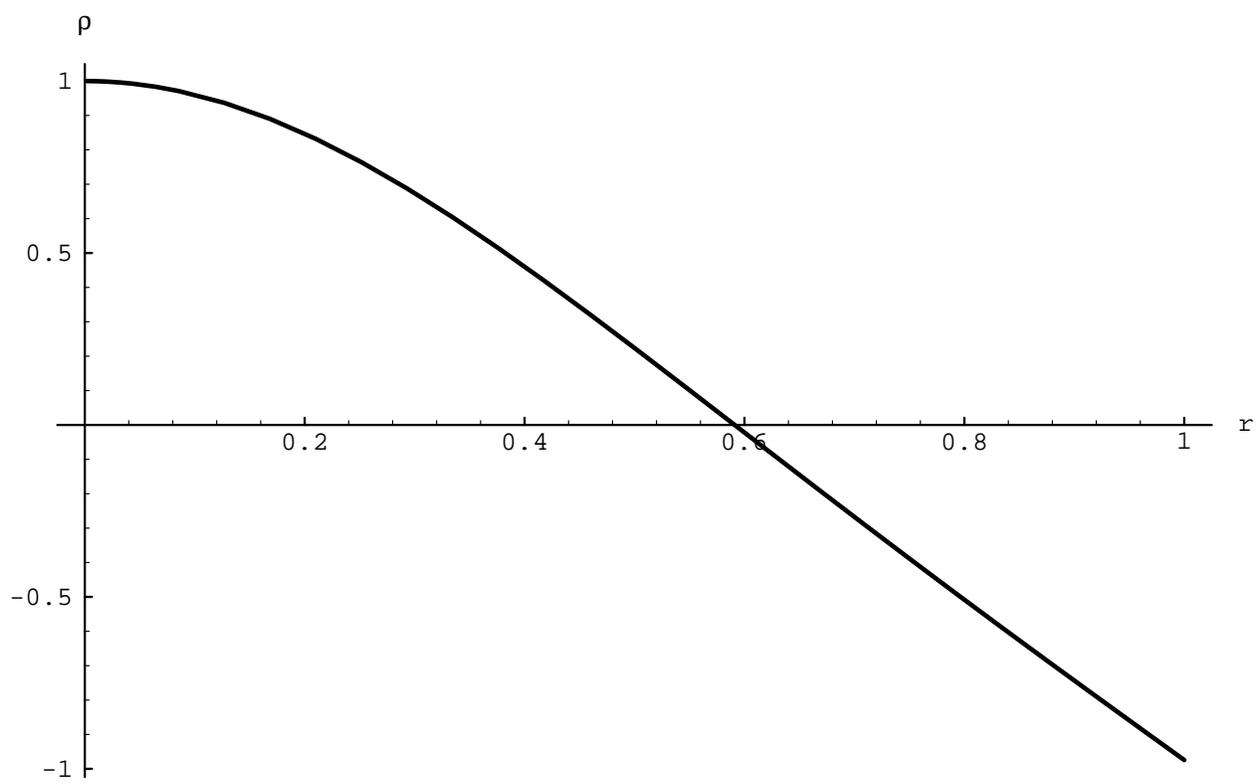
\end{document}